\documentclass[preprint,aps]{revtex4}
\usepackage{graphicx,amsfonts} 
\begin{document}
\title{Radiative corrections to the Chern-Simons term at finite
temperature in the noncommutative Chern-Simons-Higgs model}
\author{L. C. T. de Brito}
\author{M. Gomes}
\author{Silvana Perez}
 \altaffiliation[Also at]{ Departamento de F\'{\i}sica, Universidade Federal do
Par\'a, Caixa Postal 479, 66075-110, Bel\'em, PA, Brazil; e-mail:silperez@ufpa.br}
\author{A. J. da Silva}
\affiliation{Instituto de F\'{\i}sica, Universidade de S\~{a}o Paulo,
 Caixa Postal 66318, 05315-970, S\~{a}o Paulo - SP, Brazil}
\email{lcbrito,mgomes,silperez,
  ajsilva@fma.if.usp.br}

\begin{abstract}
By analyzing the odd parity part of the gauge field two and three point vertex functions,
the one-loop radiative correction to the Chern-Simons coefficient is
computed in noncommutative Chern-Simons-Higgs model at zero and at
high temperature. At high temperature, we show that the static limit
of this correction is proportional to  $T$ but the first noncommutative
correction increases as $T\log T$. Our results are analytic
functions of the noncommutative parameter.

\end{abstract}
\maketitle

\section{Introduction}
    
The analysis of the radiative corrections to the Chern-Simons
coefficient has stimulated considerable interest in the recent years
\cite{dunne_1}.  These studies pointed out that the well known
nonrenormalization theorem \cite{coleman} may become invalid whenever
infrared singularities are present.  Typical of this possibility are
situations which potentially modify the long distance behavior of the
relevant models. Examples are the breakdown of some continuous
symmetry, thermal effects and the possible noncommutativity of the
underlying space.  In this work we are going to focus on the changes
in the Chern-Simons coefficient in a model where all these effects may
occur, the noncommutativeChern-Simons-Higgs model.

 The appearance of noncommutative coordinates has an old history
\cite{jackiw} but gained impetus more recently, mainly due to its
connection with string theory \cite{witten}. One peculiar aspect of
these theories is the ultraviolet/infrared (UV/IR) mixing
\cite{seiberg}, i. e., the replacement of some ultraviolet divergences by
infrared ones. The UV/IR mixing implies in the existence of infrared
singularities which, at higher orders, may ruin the perturbative
expansion.  These infrared singularities are generated even in
theories without massless fields. However, such behavior may be
ameliorated in some supersymmetric models \cite{gomes}, so that
supersymmetry seems to play decisive role in the construction of
consistent noncommutative theories. Up to one-loop, the absence of
UV/IR mixing has also been verified for the pure $U(n)$ Chern-Simons
model which actually seems to be a  theory without any quantum correction \cite{putz}.

In a noncommutative space, it will appear trigonometric factors and,
because of them, the amplitudes are separated in two parts, the planar
and nonplanar ones. Unless by phase factors which depend only on the
external momenta, the planar part of the amplitudes are proportional
to the corresponding amplitude in the commutative case. The main
effects coming from the noncommutativity of the space can be extracted
from the nonplanar contributions. These also contain phase factors
but, unlike the planar case, they depend on the loop
momenta.

Besides the noncommutativity of the space, thermal effects also modify
the long distance behavior of field theories. Aiming to understand the
changes induced by thermal effects many features of noncommutativity
at finite temperature have been examined \cite{fischler}. At finite
temperature, it is known that for small momenta the amplitudes in a
commutative model are, in general, not well behaved and this feature
is understood in terms of the new structure introduced by the
temperature, the velocity of the heat bath \cite{bedaque}. On the
other hand, in a noncommutative space, the new structure is the
$\theta^{\mu \nu}$ tensor, which measures the noncommutativity of the
space, and it also leads to an infrared nonanalytic behavior.

Another effect which induces changes in the CS coefficient is the
spontaneous breakdown of some continuos symmetry.  In the commutative
space it is know that classically, in the absence of spontaneous symmetry
breakdown and at zero temperature, the non-Abelian coefficient of the
Chern-Simons term must be quantized \cite{jack2}. This was indeed
verified first up to one-loop in \cite{pisarski} and then extended to
all orders in the SU(N) Yang-Mills Chern-Simons model
\cite{das1}. Similar results hold for noncommutative theories
\cite{quantization}. When spontaneous symmetry breakdown is at work,
the quantization condition above mentioned is violated \cite{khlebi}.

    In this work we will study the corrections to the Chern Simons
  coefficient at finite temperature arising from the one-loop
  contributions to the gauge field two and three point vertex functions, in
  the broken phase of the noncommutative Chern-Simons-Higgs model.
  Unless for some special limits, in noncommutative finite temperature
  field theory there are many difficulties to evaluate amplitudes in a closed
  form. Thus, similarly to \cite{frenkel}, we will consider a
  generalization of the hard thermal loop limit, which involves the
  noncommutative parameter.  We then prove that in the static limit
  and at high temperature the Chern-Simons term increases like $T$ and
  actually does not depend on the noncommutative parameter. However, at
  the next level of approximation, which is linear in the noncommutative
  parameter, we found that the odd part of the two point vertex
  function increases as $T\log T$. As expected, although we are
  considering the Abelian model, because of the noncommutativity,
  there will be strong resemblances with the non-Abelian theory.

The work is organized as follows: In Section \ref{sec2} the
noncommutative version of the Chern-Simons-Higgs model in the broken
phase is presented. Section \ref{sec3} contains the one-loop
corrections to the odd parity part of the gauge field two point vertex
function, both in commutative as well as in noncommutative cases and at
finite temperature in the imaginary time formalism. The zero
temperature results are obtained as consequence of this
evaluation. The odd parity part of the gauge field three point vertex
function is studied in Section \ref{sec3a}.  Finally, in Section
\ref{sec4} the conclusions are presented. One Appendix collects some
useful integrals used in the work.

\section{Noncommutative Chern-Simons-Higgs model}
\label{sec2}

Noncommutative quantum field theories are defined in a space where
the coordinates do not commute among themselves. Rather, the
commutator between two position operators is postulated to be

\begin{equation}
[\hat x^{\mu},\, \hat x^{\nu}] = i \theta^{\mu \nu},
\end{equation}

\noindent where $\theta^{\mu \nu}$ is an antisymmetric matrix, which
for simplicity we take as commuting with the $\hat x$'s. The algebra of
operators in such space  has  been extensively studied
\cite{gronewold}, and many properties are known (see \cite{douglas}
for some reviews). A basic result is that, due to the Wigner-Moyal correspondence,
instead of working with functions of the noncommutative
coordinates, one may use ordinary functions of commutative variables
embodied with the so called Moyal product, defined as

\begin{equation}
f(x) \ast g(x) = \left[e^{(i/2) \theta^{\mu \nu}
\partial^{(\zeta)}_{\mu} \partial^{(\eta)}_{\nu}} f(x + \zeta) g(x +
\eta) \right]_{\zeta=0=\eta}.
\end{equation}

Using this definition, one can   study quantum field theories in
 a noncommutative space, by replacing the standard pointwise product 
of fields by the Moyal one. For simplicity, in this
work we shall keep $\theta^{0i}=0$.

 In the present work we will study the Chern-Simons-Higgs model in a noncommutative
space. The model is defined by the action

\begin{eqnarray}\label{lagrangian}
S &=& \frac{1}{2}\int d^{3}x \epsilon^{\mu \nu \lambda} \left[A_{\mu}
\ast \partial_{\nu} A_{\lambda} + \frac{2 ig}{3} A_{\mu}\ast
A_{\nu} \ast A_{\lambda} \right] \nonumber \\
&+& (D_{\mu} \Phi) \ast (D^{\mu} \Phi)^{\dag} -
\frac{\lambda}{4}\left[\Phi \ast \Phi^{\dag} - v^2\right]_{\ast}^2,
\end{eqnarray}

\noindent where $v$, $g$  and $\lambda$ are constants. $D_{\mu}\Phi$ is the
covariant derivative, defined in such way to ensure the gauge invariance
of the action. Because of the  noncommutativity, under a $U(1)$ gauge
transformation $U$ the basic fields may alternatively transform as: 

1) Fundamental representation: $\Phi\rightarrow \Phi *
   U$ and the covariant derivative is given by $D_{\mu}\Phi=\partial_{\mu}\Phi-ig\Phi * A_{\mu}$;

2) Anti-fundamental representation:  $\Phi\rightarrow  U^{-1} *
   \Phi$ and $D_{\mu}\Phi=\partial_{\mu}\Phi+igA_{\mu}*\Phi$;

3) Adjoint representation: $\Phi\rightarrow  U^{-1} * \Phi * U$ and
   $D_{\mu}\Phi=\partial_{\mu}\Phi+ig[A_{\mu},\Phi]_{*}$. Throughout
this article we will employ the notation $[ \, , \, ]_{\ast}$ and $\{ \, ,
\, \}_{\ast}$ to respectively designate  the commutator and anticommutator 
using the Moyal product.

\noindent In all theses cases 

\begin{equation}A_{\mu} \rightarrow U^{-1}*A_{\mu}*U
-\frac{1}{ig}(\partial_{\mu}U^{-1})*U.
\end{equation}

 In the adjoint representation, the  Higgs mechanism and the induction
 of new terms containing the gauge fields  due the spontaneous breakdown of the
gauge symmetry are absent. Because of
that we are going to restrict our considerations to the fundamental
representation (the analysis of the anti-fundamental representation is
actually very similar).

In the spontaneously broken phase, $\langle \Phi \rangle \equiv
v\ne 0$, one can choose the decomposition $\Phi = v +
\frac{1}{\sqrt{2}}(\sigma + i \chi)$ and rewrite
Eq. (\ref{lagrangian}) as

\begin{eqnarray}\label{lagrang2}
S &=&\int d^{3}x \frac{1}{2} \epsilon^{\mu \nu \lambda} \left(A_{\mu}
\ast \partial_{\nu} A_{\lambda} + \frac{2 ig}{3} A_{\mu}\ast
A_{\nu} \ast A_{\lambda} \right) + \frac{m}{2} A_{\mu} \ast
A^{\mu} - \frac{1}{2 \xi} (\partial_{\mu}
A^{\mu}){\ast}(\partial_{\nu} A^{\nu})  \nonumber \\ 
&+& \frac{1}{2}(\partial_{\mu}{\sigma})\ast(\partial^{\mu}{\sigma}) -
\frac{m_{\sigma}^2}{2} \sigma \ast \sigma +
\frac{1}{2}(\partial_{\mu}{\chi})\ast(\partial^{\mu}{\chi}) -
\frac{m_{\chi}^2}{2} \chi\ast \chi \nonumber \\ 
&-&  \frac{g}{2} A_{\mu} \ast \left(\sigma \ast  
{\buildrel\,\leftrightarrow\over{\partial^{\mu}}\!\!}
\, \, \chi  - \chi \ast  
{\buildrel\,\leftrightarrow\over{\partial^{\mu}}\!\!} \,\,
\sigma  - i [\sigma ,\partial^{\mu}\sigma]_{\ast} - i [\chi
,\partial^{\mu}\,\chi]_{\ast} \right) \nonumber \\ 
&+& \frac{g^{2}}{2}  A_{\mu} \ast A^{\mu} \ast\left(\sigma \ast \sigma
+ \chi \ast \chi + 2 \sqrt{2} v
\sigma + i [\sigma,\chi]_{\ast} \right) \nonumber \\ 
&-& \frac{\lambda}{2 \sqrt{2}} v \left\{ \sigma,
\frac{(\sigma \ast \sigma + \chi \ast \chi)}{2} +
\frac{i}{2}[\chi,\sigma]_{\ast}\right\}_{\ast} -
\frac{\lambda}{16}\left(\sigma \ast \sigma + \chi \ast \sigma + i
[\chi,\sigma]_{\ast}\right)_{\ast}^2,
\end{eqnarray}

\noindent where we have chosen the $R_{\xi}$ gauge, specified by the gauge
fixing action

\begin{equation}
S_{GF} = - \frac{1}{2\xi} \int d^{3}x\left(\partial_{\mu} A^{\mu}
+ \xi \sqrt{2}v \chi\right)_{\ast}^2,
\end{equation}

\noindent which has the merit of canceling the nondiagonal terms in
the quadratic part of the model. We have also defined

\begin{equation}
m = 2 (gv)^2, \qquad m_{\sigma}^2= \lambda v^2, \qquad m_{\chi}^2=
\xi m.
\end{equation}

 To complete the action of the model, one has to add to
  Eq. (\ref{lagrang2}) the Faddeev-Popov  action
  given by

\begin{equation}
S_{FP} = \int d^{3}x \left[ \partial_{\mu}\bar{c}*\partial^{\mu}c +
i \partial_{\mu}\bar{c}\ast\left( c \ast A_{\mu}    -
A_{\mu} \ast  c  \right) +i\xi v \bar{c}*\chi*c\right],
\end{equation}

\noindent where $\bar{c}$ and $c$ are the ghost fields.

\begin{figure*}
\includegraphics{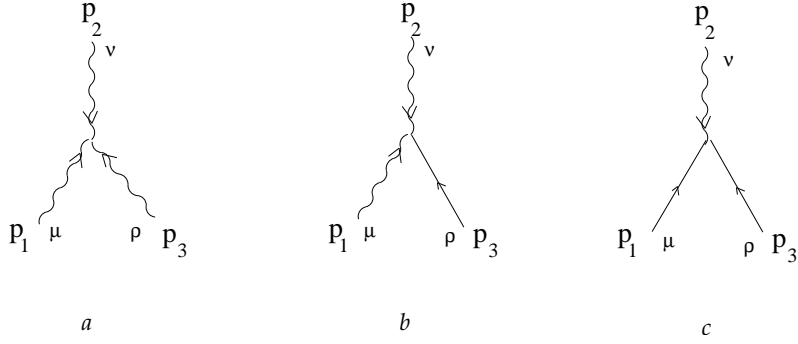}
\caption{Vertices contributing to the odd parity violating part of the
$A_\mu$ two and three point vertex function.} 
\label{Fig3}
\end{figure*}

The propagator for the gauge field is 

\begin{equation}
D_{\mu \nu} (p) = \frac{i}{p^2-m^2} \left[ - m g_{\mu \nu}
+ p_{\mu} p_{\nu} \frac{m - \xi}{p^2 + \xi m} + i
\epsilon_{\mu \nu \lambda} p^{\lambda} \right] 
\end{equation}

\noindent 
 and for the other  fields they 
 are the standard ones ($D_{\sigma}(p)=
i/(p^2 - m_{\sigma}^2)$, $D_{\chi} = i/(p^2- m_{\chi}^2)$  and 
$D_{c}={i}/{p^2}$). These propagators are not affected by the 
noncommutativity.
We will use the following analytic expression
for the vertices in the figure \ref{Fig3} (we list only those that will 
contribute in our calculation)
\begin{eqnarray}
& &iA_{\mu}*A_{\nu}*A_{\rho} \qquad  {\rm{vertex}} \qquad  \leftrightarrow \qquad
2ig\epsilon_{\mu\rho\nu}\sin(p_1\wedge p_2) 
\label{vert1}\\
&& A_{\mu}*A_{\nu}*\sigma \qquad \quad {\rm{vertex}} \qquad  \leftrightarrow \qquad
2\sqrt{2}iv g^2 g_{\mu\nu} cos(p_1\wedge p_2)
\label{vert2}\\
&& iA_\rho*[\sigma,\partial^{\rho}\sigma]_{*} \qquad {\rm{vertex}}\qquad
\leftrightarrow \qquad 2 g p^{\rho}_{3}\sin(p_2\wedge p_3)
\label{vert3}
\end{eqnarray}

At finite temperature and using the imaginary time formalism, the
gauge propagator is 

\begin{equation}
D_{\mu \nu} (p) =  \frac{1}{p^2+m^2} \left[ m
\delta_{\mu \nu} - p_{\mu} p_{\nu} \frac{m - \xi}{p^2 + \xi
m} -  \epsilon_{\mu \nu \lambda} p^{\lambda} \right],
\end{equation}
\noindent with $p^{\mu}\equiv (p^0,\vec{p})=(2 \pi n T, \vec{p})$.

\section{Two point function}
\label{sec3}

In this section we will compute the one-loop radiative correction to
the gauge field two point vertex function
of the above model at finite temperature,
in both commutative and noncommutative cases, employing   the imaginary
time formalism. In the noncommutative situation, to fix the behavior of the Chern-Simons term  we will have to examine
the corrections to the gauge field  two and three point vertex  functions. We begin by first
considering the corrections to the two point vertex function.

\subsection{Commutative case}

 Let us start by evaluating the parity violating part of the
polarization tensor $\Pi_{\mu \nu}$ in the commutative case. There is
only one diagram,  Fig. \ref{Fig1}$a$, to evaluate.  At one loop, the ghost
field does not contribute to the parity violating part,
 as can be seen from ${S}_{FP}$. Thus, we have

\begin{equation}\label{eq.01}
\Pi_{\mu \nu}^{odd} (p)\equiv {\cal{\pi}}_{\mu \nu} (p) =8 (v\,g^2)^2 T \sum_{n} 
\int \frac{d^2k}{(2 \pi)^2} \frac{\epsilon_{\mu \nu\lambda} k^{\lambda}}{(k^2 +
m^2)[(p-k)^2 + m^2_{\sigma}]}. 
\end{equation}

\noindent  The consideration of the static
limit, where $p_0=0$ and $\mid \vec{p} \mid$ is small, yields

\begin{equation}\label{tensorI}
{\cal{\pi}}_{0i}(p_0=0)= 8 (v\,g^2)^2\int
\frac{d^2k}{(2 \pi)^2} \epsilon_{0ij} k^j T \sum_{n}\frac{1}{k_0^2 +
w_m^2} \frac{1}{k_0^2 + w_{\sigma}^2(p)}.
\end{equation}

\noindent Note that, in the static limit, ${\cal{\pi}}_{ij}$ vanishes
as the integrand is an odd function of $n$. We have also introduced
the notation $w_m^2=\vec{k}^2 + m^2$ and
$w_{\sigma}^2(p)=(\vec{p}-\vec{k})^2 + m_{\sigma}^2$. Using that

\begin{eqnarray}\label{expansion}
\frac{1}{k_0^2 + (\vec{p}-\vec{k})^2 + m_{\sigma}^2} &=& \frac{1}{k_0^2 +
w_{\sigma}^2} + \frac{2 \vec{k} \cdot \vec{p} }{(k_0^2 +
w_{\sigma}^2)^2} + {\cal{O}}(\vec{p}^{\phantom a 2})\nonumber \\
&=& \left(1 - 2 \vec{k} \cdot \vec{p} \frac{\partial}{\partial
m_{\sigma}^2} \right) \frac{1}{k_0^2 +
w_{\sigma}^2} + {\cal{O}}(\vec{p}^{\phantom a 2}),
\end{eqnarray}

\noindent 
where $w_\sigma\equiv w_\sigma(0)$, we write Eq.  (\ref{tensorI})
for small external momentum as

\begin{equation}
{\cal{\pi}}_{0i}(p_0=0) = 8 (v\,g^2)^2\int \frac{d^2k}{(2
\pi)^2} \left(1 - 2 \vec{k} \cdot \vec{p} \frac{\partial}{\partial
m_{\sigma}^2} \right) \epsilon_{0ij} k^j T \sum_{n}\frac{1}{k_0^2 +
w_m^2} \frac{1}{k_0^2 + w_{\sigma}^2} + {\cal{O}} (\vec{p}^{\phantom a
  2}).
\end{equation}

After that, we evaluate the sum in $k_0$ through the use of

\begin{equation}\label{sum}
\sum_{n=-\infty}^{n=+\infty} f(n) = - \pi \sum'
[f(z)\cot(\pi z)],
\end{equation}

\noindent where the prime in the sum indicates that it runs over the 
residues of the poles of $f(z)$. Thus, we find

\begin{eqnarray}\label{planar}
{\cal{\pi}}_{0i}(p_0=0) &=&  4 (v\,g^2)^2\int
\frac{d^2k}{(2 \pi)^2} \epsilon_{0ij} k^j \nonumber \\
&\times& \left(1 - 2 \vec{k} \cdot \vec{p} \frac{\partial}{\partial
m_{\sigma}^2}\right) \left\{ \frac{1}{m_{\sigma}^2 - m^2} 
\left[ \frac{\coth{(\beta w_m/2)}}{w_m} -
\frac{\coth{(\beta w_{\sigma}/2)}}{w_{\sigma}}\right] \right\} .
\end{eqnarray}

Using that $\coth{(\beta x/2)}=1 + \frac{2}{e^{\beta x}-1}$, we can separate
 the zero temperature from the finite temperature part. The zero temperature
part can be computed straightforwardly, giving

\begin{eqnarray}
{\cal{\pi}}_{0i}(p_0=0;T=0) &=& 4 (v\,g^2)^2 \int
\frac{d^2k}{(2 \pi)^2} \epsilon_{0ij} k^j  \left(1 - 2 \vec{k} \cdot \vec{p} 
\frac{\partial}{\partial m_{\sigma}^2}\right)  \nonumber \\
&\times&\left[ \left( \frac{1}{m_{\sigma}^2 - m^2}\right) 
\left( \frac{1}{w_m} -
\frac{1}{w_{\sigma}}\right) \right]= \frac{2g^2}{3 \pi}
\frac{(1 + \frac{1}{2}
\frac{m_{\sigma}}{m})}{(1 + \frac{m_{\sigma}}{m})^2}\epsilon_{0ij}p^j.
\label{planar0}
\end{eqnarray}

The finite temperature part in Eq. (\ref{planar}) is more complicated
but can be expressed in terms of polylogarithm functions as

\begin{eqnarray}
{\cal{\pi}}_{0i}(p_0=0;T)&=& - \frac{4 (v\,g^2)^2}{\pi} \epsilon_{0ij}
p^j \frac{\partial}{\partial m_{\sigma}^2} 
\left(\frac{f(m,m_{\sigma},T)}{m_{\sigma}^2 -m^2}\right) 
\end{eqnarray}

\noindent  where \cite{leon}  

\begin{equation}\label{function1}
f(m,m_{\sigma},T)=\left[ T^3\, {\rm PolyLog}(3,e^{-\beta m}) + m \,T^2\,
{\rm PolyLog}(2,e^{-\beta m})\right] - [ m\rightarrow m_{\sigma}]
\end{equation}

\noindent
and

\begin{equation}\label{poly}
{\rm PolyLog}(b, a) \equiv \sum_{n=1}^{\infty} \frac{a^n}{n^b}.
\end{equation}

In the high temperature limit, the above expression
furnishes 

\begin{equation}\label{planarT}
{\cal{\pi}}_{0i} (p_0=0;T)=-\frac{4 (v\,g^2)^2}{\pi} \epsilon_{0ij}p^j
T \frac{\partial}{\partial m_{\sigma}^2} \left[\frac{m_{\sigma}^2 \log
(m/{m_{\sigma}})}{m_{\sigma}^2 - m^2} \right].
\end{equation}

The results (\ref{planar0}) and (\ref{planarT}) agree with the
corresponding ones in the Chern-Simons-Higgs limit of the
Maxwell-Chern-Simons-Higgs model calculated in \cite{parityvio}.

\subsection{Noncommutative case}

\begin{figure*}
\includegraphics{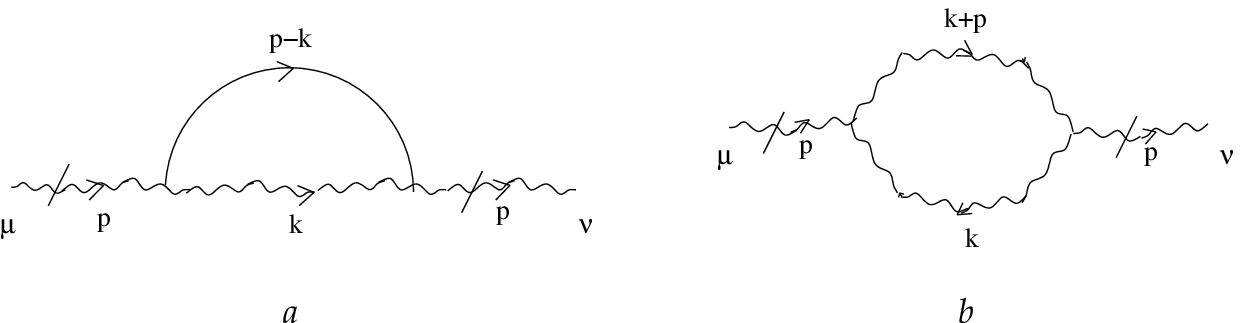}
\caption{One-loop graphs contributing to the parity violating part of the
$A_\mu$ two point vertex function.} 
\label{Fig1}
\end{figure*}

Next, let us determine the parity violating part of the polarization
tensor in the noncommutative Chern-Simons-Higgs model. In this case,
 both diagrams in Fig. \ref{Fig1} contribute. As in the commutative case,
at one-loop the ghost field does not give a parity violating part
correction to the polarization tensor. The diagram in Fig.  \ref{Fig1}$a$  gives

\begin{equation}
\Pi^{odd}_{a,{\mu\nu}} \equiv {\cal{\pi}}_{a,\mu \nu}
(p)=8 (v\,g^2)^2T \sum_{n} \int \frac{d^2 k}{(2 \pi)^2}
\frac{\epsilon_{\mu \nu\lambda} k^{\lambda} \cos^2(k \wedge p)}{(k^2 +
m^2)[(p-k)^2 + m^2_{\sigma}]},
\end{equation}
 
\noindent where $k \wedge p = \frac{1}{2} \theta^{\mu \nu} k_{\mu}
p_{\nu}=\frac12 k_\mu {\tilde p}^\mu$  with ${\tilde p}^\mu\equiv 
\theta^{\mu\nu}p_\nu$.  Again, we will consider only the static limit. Therefore,
we have

\begin{eqnarray}
{\cal{\pi}}_{a,0i}(p_0=0) &=& 4 (v\,g^2)^2\int
\frac{d^2k}{(2 \pi)^2} \left[ 1 + \cos(2 \vec{k} \wedge \vec{p})
\right] \epsilon_{0ij} k^j \nonumber \\  &\times& T
\sum_{n}\frac{1}{k_0^2 + w_m^2} \frac{1}{k_0^2 + w_{\sigma}^2(p)}.
\end{eqnarray}

By following the same steps as in the calculation presented in the
previous subsection we arrive at 

\begin{eqnarray}
{\cal{\pi}}_{a,0i} (p_0=0) &=& 2 (v\,g^2)^2\int
\frac{d^2k}{(2 \pi)^2} \left[1 + \cos(\vec{k} \cdot \tilde{\vec{p}})
\right] \epsilon_{0ij} k^j \nonumber \\ 
&\times& \left(1 - 2 \vec{k}
\cdot \vec{p} \frac{\partial}{\partial m_{\sigma}^2} \right) \left\{
\frac{1}{m_{\sigma}^2 - m^2} \left[\frac{\coth{(\beta w_m/2)}}{w_m} -
\frac{\coth(\beta w_{\sigma}/2)}{w_{\sigma}} \right] \right\} \nonumber
\\ &\equiv& A_{0i} + B_{0i}\label{noncommut}
\end{eqnarray}

\noindent where $A_{0i}$ and $B_{0i}$ are respectively the planar and
nonplanar contributions to ${\cal{\pi}}_{{a},0i}(p_0=0)$; the
$\tilde{p}\equiv\mid \vec{\tilde{p}} \mid\rightarrow 0$ limit will be
taken shortly.  The planar part is exactly one half of
Eq. (\ref{planar0}). Thus, considering the nonplanar contribution,
from Eq. (\ref{noncommut}) we have

\begin{eqnarray}
B_{0i} &=& 2 (v\,g^2)^2 \int \frac{d^2k}{(2 \pi)^2} 
\cos( \vec{k} \cdot \tilde{\vec{p}}) \epsilon_{0ij} k^j \nonumber \\
 &\times & \left(1 - 2 \vec{k} \cdot \vec{p} \frac{\partial}{\partial
 m_{\sigma}^2} \right) \left\{ \frac{1}{m_{\sigma}^2 - m^2}
 \left[\frac{\coth{(\beta w_m/2)}}{w_m} - \frac{\coth(\beta
 w_{\sigma}/2)}{w_{\sigma}} \right] \right\} \nonumber \\  &=& -
 4 (v\,g^2)^2 \int \frac{d^2k}{(2 \pi)^2} \cos(\vec{k} \cdot
 \vec{\tilde{p}}) \epsilon_{0ij} k^j \vec{k} \cdot \vec{p}\nonumber
 \\  &\times& \frac{\partial}{\partial m_{\sigma}^2} \left\{
 \frac{1}{m_{\sigma}^2 - m^2}\left[\frac{\coth{(\beta w_m/2)}}{w_m} -
 \frac{\coth(\beta w_{\sigma}/2)}{w_{\sigma}} \right] \right\}
 \nonumber \\  &=& - 4 (v\,g^2)^2 {\epsilon_{0ij} p^j}
  \int\frac{ d^2 k}{(2 \pi)^2} \frac{\cos(\vec{k} \cdot \vec{\tilde{p}})
 (\vec{k} \cdot \vec{p})^2}{|\vec{p}|^2} \nonumber \\  &\times&
 \frac{\partial}{\partial m_{\sigma}^2} \left\{ \frac{1}{m_{\sigma}^2
 - m^2} \left[\frac{\coth{(\beta w_m/2)}}{w_m} - \frac{\coth(\beta
 w_{\sigma}/2)}{w_{\sigma}} \right] \right\},
\end{eqnarray}

\noindent where we have used polar coordinates such that

\begin{equation}
\vec{k}\equiv \frac{(\vec{k} \cdot \vec{p}) \vec{p}}{|\vec{p}|^2} +
\frac{(\vec{k} \cdot \vec{\tilde{p}})
\vec{\tilde{p}}}{|\vec{\tilde{p}}|^2}.
\end{equation}
 
The angular part of the above integral  can be expressed in terms of
Bessel functions, as follows

\begin{equation}\label{integral}
B_{0i} = - 2 (v\,g^2)^2 \frac{\epsilon_{0ij} p^j}{\pi}
\frac{\partial}{\partial m_{\sigma}^2} \left\{ \frac{1}{m_{\sigma}^2 -
m^2}\int_0^{\infty} dk \frac{k^2}{\tilde p} J_1 (k \tilde p )
\left[\frac{\coth{(\beta w_m/2)}}{w_m} - \frac{\coth(\beta
w_{\sigma}/2)}{w_{\sigma}} \right]\right\}.
\end{equation}

\noindent 
These integrals
 are evaluated in the appendix \ref{a1}. Here, we will only write the
 final results. So, we have:

\begin{equation}\label{nonplanar0}
B_{0i}(T=0)= - \frac{2 (v\,g^2)^2}{ \pi} \epsilon_{0ij} p^j
\frac{\partial}{\partial m_{\sigma}^2} \left\{ \frac{1}{m_{\sigma}^2 -
m^2} \left[\left(m \frac{e^{- \tilde p m}}{(\tilde p )^2} + \frac{e^{-
\tilde p m}}{(\tilde p )^3}\right) - (m \rightarrow m_{\sigma}
)\right] \right\}
\end{equation}

\noindent and

\begin{eqnarray}\label{nonplanarT}
B_{0i}(T)&=&- \frac{4 (v\,g^2)^2}{\pi} \epsilon_{0ij} p^j 
\frac{\partial}{\partial m_{\sigma}^2} \left\{  \frac{1}{m_{\sigma}^2
- m^2}\right. \nonumber \\ 
&\times&\left. \left[ \left(m T^2 \sum_{n=1}^{\infty} \frac{e^{- \beta m
\sqrt{n^2+\tau^2}}} {n^2 + \tau^2} + T^3 \sum_{n=1}^{\infty}
\frac{e^{- \beta m \sqrt{n^2+\tau^2}}} {(n^2 + \tau^2)^{3/2}}\right) -
\left( m \rightarrow m_{\sigma}\right) \right] \right\}, 
\end{eqnarray} 

\noindent where $\tau\equiv\tilde p T$. Note the apparent singularity
in Eq. (\ref{nonplanar0}) at $\tilde{p} =0$; this kind of
structure is the well known infrared singularity, characteristic of
noncommutative field theories \cite{seiberg}.  Here, however, as
 we will shortly see, this singularity cancels in the final
result. Next, let us check if these results are consistent with the
 $\theta=0$ limit, namely, if in this limit we obtain the 
other half of Eqs. (\ref{planar0}) and (\ref{planarT}), so that the
 commutative result is recovered. For the zero temperature part,
expanding Eq. (\ref{nonplanar0}) for small values of $\tilde{p} $, we
 get

\begin{equation}\label{eq.03}
B_{0i}(T=0)= \epsilon_{0ij} p^j \left[ \frac{g^2}{3 \pi}
\frac{(1 + \frac{1}{2}
\frac{m_{\sigma}}{m})}{(1 + \frac{m_{\sigma}}{m})^2} - \frac{(v\,g^2)^2}{4
\pi} \tilde p  \right] + {\cal{O}}(\tilde{p}^{\phantom a 2}).
\end{equation}

\noindent When $\theta \rightarrow 0$, we obtain one half of
Eq.$\!$ (\ref{planar0}), completing the expected result for the
commutative case.

Now, looking at the temperature dependent part,
Eq. (\ref{nonplanarT}), when $\theta=0$, we found that the result is
again proportional to the function defined in
Eq. (\ref{function1}):

\begin{eqnarray}\label{eq.02}
B_{0i}(T) &=&  - \frac{4 (v\,g^2)^2}{\pi} \epsilon_{0ij} p^j
 \frac{\partial}{\partial m_{\sigma}^2} 
\left(\frac{f(m,m_{\sigma},T)}{m_{\sigma}^2 - m^2}\right).\end{eqnarray} 

From the asymptotic behavior of the polylogarithm functions
\cite{leon}, we found that in the high temperature limit this gives

\begin{equation}
B_{0i}(T) \rightarrow \frac{-2 (v\,g^2)^2}{\pi} \epsilon_{0ij} p^j T
 \frac{\partial}{\partial m_{\sigma}^2}\left[\frac{m_{\sigma}^2
 \log(m/m_{\sigma})}{m_{\sigma}^2 - m^2}\right] \equiv \frac{1}{2}
 {\cal{\pi}}_{0i} (p_0=0,T).
\end{equation}

Once we have obtained the commutative limit, we consider next the first
correction,  which in this case is proportional to $\tau^2$. The
computations for this term are similar to the ones that we have done 
 so far, and finally, we can write the result, up to $\tau^2$, as

\begin{eqnarray}\label{eq.04}
B_{0i}(T)&=&
- \frac{2 (v\,g^2)^2}{\pi} \epsilon_{0ij} p^j  T 
\frac{\partial}{\partial m_{\sigma}^2}\left\{ \frac{1}{m_{\sigma}^2 -
m^2}\right. \nonumber \\ 
&\times& \left. \left[ m^2_{\sigma} \log(m/m_{\sigma}) + \tau^2\frac{1
}{8 T^2}\left(m^4 \log(m/T) - m_{\sigma}^4 \log(m_{\sigma}/T)
\right)\right] \right\} + {\cal{O}} (\tau^4).
\end{eqnarray}
 
Before proceeding to the computation of the remaining diagram, it is
worth noting that both the temperature independent as well as
the temperature dependent parts of $B_{0i}$, Eqs. (\ref{eq.03}) and
(\ref{eq.04}), are analytic functions of $\tilde{p}$: no infrared
singularities shows up.

Evaluating the  graph in Fig. \ref{Fig1}$b$, we have

\begin{equation}
\Pi_{b,\mu \nu} (p)=2 g^2T \sum_n \int \frac{d^3
k}{(2 \pi)^2} \epsilon_{\mu \alpha \beta} \epsilon_{\sigma \rho \nu}
\sin^2(k \wedge p) D^{\alpha \rho} (k+p) D^{\sigma \beta} (k)
\end{equation}

\noindent and considering only the parity violating part of this diagram,
we obtain

\begin{eqnarray}
{\cal{\pi}}_{b,\mu \nu}(p) &=& 2 g^2T \sum_n \int \frac{d^3
k}{(2 \pi)^3} \frac{\sin^2(k \wedge p) \epsilon_{\mu \nu \lambda}}{(k^2+m^2
) [(k+p)^2+m^2]} \nonumber \\
&\times& \left\{m p^{\lambda} -  (m - \xi) k
\cdot (k+p) \left[ \frac{k^{\lambda}}{k^2 + \xi m} -
\frac{(k+p)^{\lambda}}{(k+p)^2 + \xi m} \right]\right\}.
\end{eqnarray}

The integrals that appear in this expression are similar to the
ones we have computed before. So, without going into details, for
  $\vec p \rightarrow 0$ after taking $p_0=0$   the
calculation gives

\begin{equation}
{\cal{\pi}}_{b, 0i} (p_0=0;T=0) = \frac{g^2}{16 \pi}(3
m- \xi ) \tilde{p}\, \epsilon_{0ij} p^j
\end{equation}

\noindent so that at $T=0$ we obtain the radiative correction to the
odd part of the two point function 

\begin{equation}
\Pi^{odd}_{0i}(T=0)=\left[\frac{2g^2}{3 \pi}
\frac{(1 + \frac{1}{2}
\frac{m_{\sigma}}{m})}{(1 + \frac{m_{\sigma}}{m})^2}
+\frac{ g^2 \tilde{p}}{16\pi}(m-\xi)\right]\epsilon_{0ij} p^j.
\label{ref39}
\end{equation}

\noindent
Actually, the above expression with the replacement of $\epsilon_{0ij} p^j$
by $\epsilon_{\mu\nu\rho} p^\rho$ holds also for $\pi_{\mu\nu}$.

On the other hand,  for  high temperature and also small
$\vec p$ and $\tau$  we have
\begin{equation}
{\cal{\pi}}_{b, 0i} (p_0=0;T) = -
\frac{\epsilon_{0ij}p^j}{16 \pi} \frac{g^2\tau^2}{T}\left[2 m
\log(m/T) + (m - \xi)F\right]
\end{equation}

\noindent where

\begin{equation}
F\equiv \frac{\xi^2 (\xi  + m)}{(\xi  -
m)^3}\log(\sqrt{\xi m}/T) - \frac{m(m^2 + 4 \xi^2  - 3 \xi
m)}{(\xi  - m)^3} \log(m/T).
\end{equation}

\noindent The complete result for the two point  function at high $T$  in this limit is

\begin{eqnarray}\label{resT}
{\cal{\pi}}_{0i}^{NC} (p_0=0;T) &=& - \frac{1}{\pi}
\epsilon_{0ij} p^j T \nonumber \\ 
&\times& \left\{ 2 (ve^2)^2 \frac{\partial}{\partial m_{\sigma}^2}\left[
\frac{2m^2_{\sigma}\log(m/m_{\sigma}) + \frac{\tilde{p}^2}{8}(m^4
\log(m/T) - m^4_{\sigma}\log(m_{\sigma}/T))}{m_{\sigma}^2 - m^2}\right]
\nonumber \right. \\
&+& \left. \frac{g^2\tilde{p}^2}{16} \left[2 m \log(m/T) + (m - \xi) F\right]
  \right\} .
\end{eqnarray} 

It is worth noting here the gauge dependence of
${\cal{\pi}}_{b}$. Naively, we could expect that there would be no
gauge dependence at this point of the calculation, as it happens
in the commutative
case. But, this graph gives a purely noncommutative contribution to
$\Pi_{\mu \nu}$, which will disappear when $\theta$ goes to
zero. Therefore, there is no relation between it and what
is found in the commutative case. We recall that a dependence on the gauge
parameter was also  obtained in the commutative studies of the
non-Abelian gauge models \cite{pisarski}.

\section{Three point function}
\label{sec3a}

To complete our analysis on the radiative corrections to the
Chern-Simons term we will now compute the odd parity part of the
one-loop contribution to the gauge field three-point vertex function.  
For simplicity, we shall calculate only the small momenta leading
corrections. As we will see, this implies that the contributions in which we
are interested are proportional to the product of the Levi-Civit\`a symbol by a
trigonometric sine factor.

To extract the $T=0$ contribution we will use  the identity  \cite{kapusta}

\begin{eqnarray}
\frac{i}{\beta}\sum_{n}f(\vec{k},k_0=\frac{2\pi n}{\beta})&=&
\int_{-i\infty+\epsilon}^{i\infty+\epsilon}dk_{0}
\frac{f(\vec{k},k_{0})}{e^{\beta k_{0}}-1}+\int_{-i\infty-\epsilon}^{i\infty-\epsilon}dk_{0}\frac{f(\vec{k},k_{0})}{e^{-\beta
k_{0}}-1}\nonumber\\ &+&\int_{-i\infty}^{i\infty}dk_{0}f(\vec{k},k_{0})
\label{ComplexInt}
\end{eqnarray}

\noindent
and employ the more compact notation 
\begin{eqnarray}
\int_{T=0}\frac{d^{3}k}{(2\pi)^{3}}&\equiv&\int\frac{d^{2}k}{(2\pi)^{2}}
\int_{-i\infty}^{i\infty}\frac{dk_{0}}{(2\pi)},\label{zerot}\\
\int_{T\neq0}\frac{d^{3}k}{(2\pi)^{3}}&\equiv&\int\frac{d^{2}k}{(2\pi)^2}
\left[\int_{-i\infty+\epsilon}^{i\infty+\epsilon}\frac{dk_{0}}{(2\pi)}
\frac{1}{e^{\beta
k_{0}}-1}+\int_{-i\infty-\epsilon}^{i\infty-\epsilon}\frac{dk_{0}}{(2\pi)}
\frac{1}{e^{-\beta k_{0}}-1}\right].\label{nzerot}
\end{eqnarray}  

  The relevant graphs are drawn in Fig. \ref{Fig2}.
 Here we will present our results for zero and nonzero
temperature.

\subsection{$T=0$ results} 
\begin{figure*}
\includegraphics{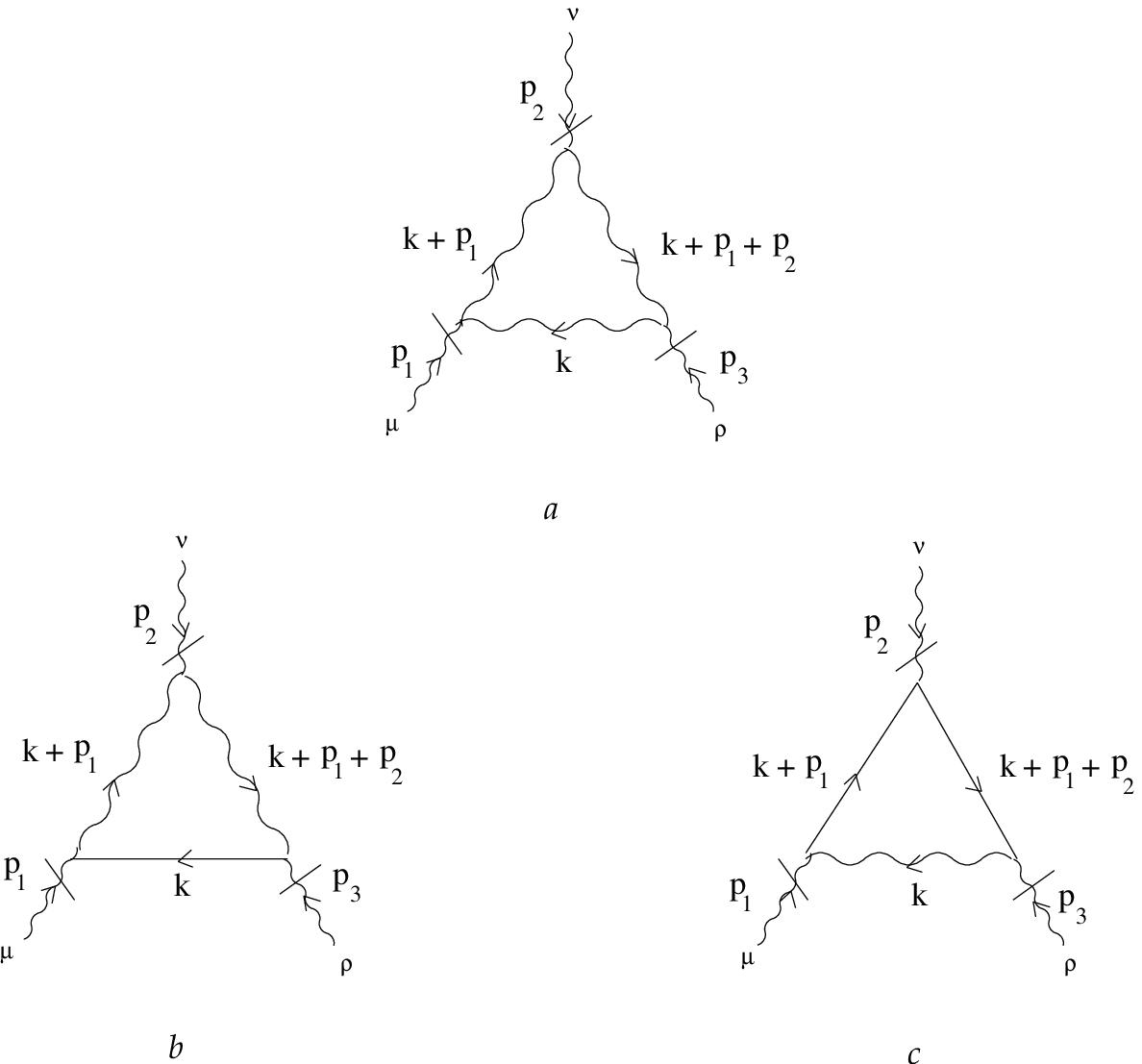}
\caption{One-loop graphs contributing to the parity violating part of the
$A_\mu$ three point vertex function.} 
\label{Fig2}
\end{figure*}

Using the expressions for the propagators and vertices listed in the
previous section, the amplitudes for the graphs which contribute to
the odd parity part of three point vertex function of the gauge field  are:

(1) Graph in Fig. \ref{Fig2}$a$

\begin{equation}
\Gamma^{\ref{Fig2}a}_{\mu \nu \rho }=-8g^3\int_{T=0} \frac{d^{3}k}{(2\pi
  )^{3}}\mathcal{D}^{\sigma \alpha }(k+p_{1})\mathcal{D}^{\beta \tau
}(k-p_{3})\mathcal{D}^{\xi \lambda }(k)\epsilon _{ \alpha\nu\beta
}\epsilon _{\lambda \mu \sigma  }\epsilon _{\tau \rho  \xi
  }S_{1},\label{eq:AnalitGraf1}\end{equation}

\noindent where
\begin{eqnarray}
S_{1}&=&\sin(k\wedge p_1)\sin(k\wedge p_3)\sin[(k+p_1)\wedge p_2]\label{S1}\\
&=&-\frac{1}{4}\Big\{\sin(p_{1}\wedge p_{2})+\sin[(2k+p_{2})\wedge p_{1}]+
\sin[(2k+p_{1})\wedge p_{2}]+\sin[(2k+p_{1})\wedge p_{3}]\Big\}.\nonumber
\end{eqnarray}

(2) Graph in Fig. \ref{Fig2}$b$

\begin{equation}
\Gamma _{\mu \nu \rho }^{\ref{Fig2}b}=\frac{-112iv^2g^5}{3}\int_{T=0} 
\frac{d^{3}k}{(2\pi
  )^{3}}\mathcal{D}_{\mu \alpha }(k+p_{1})\mathcal{D}_{\beta \rho
}(k-p_{3})\Delta _{\sigma }(k)\epsilon _{\nu }^{\, \, \, \beta \alpha
  }S_{2},\label{eq:AnalitGraf2}\end{equation}

\noindent where
\begin{eqnarray}
S_{2}&=&\cos(k\wedge p_1)\cos(k\wedge p_3)\sin[(k+p_1)\wedge p_2]\label{eq:IdentSenos}\\
&=&\frac{1}{4}\Big\{\sin(p_{1}\wedge
  p_{2})-\sin[(2k+p_{2})\wedge
  p_{1}]+\sin[(2k+p_{1})\wedge
  p_{2}]-\sin[(2k+p_{1})\wedge p_{3}]\Big\}.\nonumber
\end{eqnarray}

(3) Graph in Fig. \ref{Fig2}$c$

\begin{equation}
\Gamma _{\mu \nu \rho }^{\ref{Fig2}c}=-80v^2g^5\int_{T=0} \frac{d^{3}k}{(2\pi )^{3}}(p_{2}+k)_{\nu }\mathcal{D}_{\rho\mu}(k)\Delta _{\sigma}(k+p_{1})\Delta _{\sigma}(k-p_{3})S_{3},\label{eq:AnalitGraf3}\end{equation}

\noindent where
\begin{eqnarray}
S_{3}&=&\cos(k\wedge p_{2})\cos(k\wedge p_{3})\sin[(k+p_{2})\wedge p_{1}]\label{S3}\\
&=&\frac{1}{4}\Big\{\sin(p_{2}\wedge
p_{1})-\sin[(2k+p_{1})\wedge p_{2}]+\sin[(2k+p_{2})\wedge p_{1}]-\sin[(2k+p_{2})\wedge p_{3}]\Big\}.\nonumber
\end{eqnarray}

In the planar parts, which are the integrals containing the first term
in $S_1$, $S_2$ and $S_3$, we separate the trigonometric factor and
set equal to zero the external momenta in the propagators. In the
remaining (nonplanar) parts we approximate the sines by their
arguments and then perform the integrals. 

It turns out that  within our approximation,
$\Gamma^{\ref{Fig2}a}_{\mu \nu \rho }$ does not possess an odd
parity part. In fact, because of momentum conservation it is readily seen
that $S_1$ becomes null if the sines are replace by their arguments.
We then consider separately the odd parity part of the
other contributions. We have

\begin{eqnarray}
\left[\Gamma^{\ref{Fig2}b}_{\mu \nu \rho }\right]_{odd}&=&\frac{-112v^2g^5}{3}\int_{T=0} \frac{d^{3}k}{(2\pi)^{3}}\frac{S_{2}}{\left[(k+p_{1})^{2}-m^{2}\right]\left[(k-p_{3})^{2}-m^{2}\right]\left[k^{2}-m_{\sigma }^{2}\right]}\nonumber\\
&\times& \left \{-m^{2}\epsilon_{\mu \rho \nu }+m^{2}\left[\frac{\epsilon_{\rho \nu
\alpha }k^{\alpha }k_{\mu }}{\left[(k+p_{1})^{2}+\xi
m\right]}-\frac{\epsilon_{\mu \nu \alpha }k^{\alpha }k_{\rho
}}{\left[(k-p_{3})^{2}+\xi m\right]}\right]-\epsilon_{\mu \rho \alpha
  }k^{\alpha}k_{\nu }\right \}\label{Graph2odd},
\end{eqnarray}

\noindent
for the graph \ref{Fig2}$b$ and

\begin{equation}
\left[\Gamma^{\ref{Fig2}c}_{\mu \nu \rho }\right]_{odd}=-80v^2g^5\int_{T=0} 
\frac{d^{3}k}{(2\pi
)^{3}}\frac{\epsilon_{\rho\alpha\mu}k^{\alpha}k_{\nu}}{\left
[(k+p_1)^{2}-m^{2}_{\sigma }\right]\left[(k-p_{3})^{2}-m^{2}_{\sigma }\right]
\left[k^{2}-m^{2}
\right]}S_{3}.
\label{Graph3odd}\end{equation}

\noindent
for the graph \ref{Fig2}$c$.
Proceeding now in the way above indicated we get

\begin{equation}
\left[\Gamma^{\ref{Fig2}b}_{\mu \nu \rho}
\right]_{odd}=\frac{-112v^2g^5}{3}\left\{ \left[\Gamma^{\ref{Fig2}b}_{\mu\nu
\rho}\right]_{odd}^{planar} +\left[X^{\ref{Fig2}b}_{\mu \nu \rho }\right]_{odd} \right\},
\label{Graph2odd2}
\end{equation}

\noindent where 
\begin{eqnarray}
\left[\Gamma^{\ref{Fig2}b}_{\mu \nu \rho}\right]_{odd}^{planar}&=& \sin(p_1\wedge p_2)\int_{T=0} \frac{d^{3}k}{(2\pi
)^{3}}\frac{1}{\left[k^{2}-m^{2}\right]^2\left[k^{2}-m_{\sigma }^{2}\right]}\nonumber\\
&\times& \left\{-m^{2}\epsilon_{\mu \rho \nu }+m^{2}\left[\frac{\epsilon_{\rho \nu
\alpha }k^{\alpha }k_{\mu }}{\left[k^{2}+\xi
m\right]}-\frac{\epsilon_{\mu \nu \alpha }k^{\alpha }k_{\rho
}}{\left[k^{2}+\xi m\right]}\right]-\epsilon_{\mu \rho \alpha
  }k^{\alpha}k_{\nu }\right\}\nonumber\\
&=&\left[\frac{ -m_{\sigma }}{12 \pi(m+m_{\sigma})^{2}}\right]\sin(p_{1}\wedge p_{2})\epsilon _{\mu \rho \nu }
\label{2bPlanar}
\end{eqnarray}

\noindent and
\begin{eqnarray}
\left[X^{\ref{Fig2}b}_{\mu \nu \rho }\right]_{odd}&=&\int_{T=0} \frac{d^{3}k}{(2\pi
)^{3}}\frac{(-\sin(2k+p_{2})\wedge
  p_{1}+\sin(2k+p_{1})\wedge
  p_{2}-\sin(2k+p_{1})\wedge p_{3})}{\left[(k+p_1)^{2}-m^{2}\right]\left[(k-p_3)^{2}-m^{2}\right]\left[k^{2}-m_{\sigma }^{2}\right]}\nonumber\\
&\times& \left\{-m^{2}\epsilon_{\mu \rho \nu }+m^{2}\left[\frac{\epsilon_{\rho \nu\alpha }k^{\alpha }k_{\mu }}{\left[(k+p_1)^{2}+\xi
m\right]}-\frac{\epsilon_{\mu \nu \alpha }k^{\alpha }k_{\rho
}}{\left[(k-p_3)^{2}+\xi m\right]}\right]-\epsilon_{\mu \rho \alpha
  }k^{\alpha}k_{\nu }\right\}\nonumber\\
&=&\left[\frac{m^2+m  m_{\sigma }+ 2 m^2_{\sigma }}{30\pi(m+m_{\sigma})^{3}}\right]\sin(p_{1}\wedge p_{2})\epsilon _{\mu \rho \nu }.
\label{2bNPlanar}
\end{eqnarray}

\noindent
By summing (\ref{2bPlanar}) and (\ref{2bNPlanar}) we obtain the total contribution  from the graph \ref{Fig2}$b$ 

\begin{equation}
\left[\Gamma^{\ref{Fig2}b}_{\mu \nu \rho
}\right]_{odd}=\frac{28ig^5v^2}{45\pi}\left[\frac{(m^2_{\sigma}-2m^2
+3mm_{\sigma})}{(m+m_{\sigma})^3}\right]\epsilon_{\mu \nu \rho}
\sin(p_1\wedge p_2).\label{11}
\end{equation}

The computation for  $\left[\Gamma^{\ref{Fig2}c}_{\mu\nu\rho }\right]_{odd}$ follows
similarly yielding

\begin{equation}
\left[\Gamma^{\ref{Fig2}c}_{\mu \nu
    \rho}\right]_{odd}=\frac{ig^5v^2}{3\pi}\left[\frac{13m^2_{\sigma}
+39mm_{\sigma} +28m^2}{(m+m_{\sigma})^3}\right]\epsilon_{\mu \nu
    \rho}\sin(p_1\wedge p_2),\label{12}
\end{equation}

\noindent
so that the complete result for one loop  three point vertex  function
correction at $T=0$ is given by

\begin{eqnarray}
\left[\Gamma_{\mu\nu\rho }\right]_{odd}&=&
\left[\Gamma^{\ref{Fig2}b}_{\mu\nu\rho }\right]_{odd}+
\left[\Gamma^{\ref{Fig2}c}_{\mu\nu\rho }\right]_{odd}
\nonumber\\
&=&-\frac{i v^2 g^5}{45\pi(m+m_{\sigma })^{3}}\sin(p_1\wedge p_2) \epsilon_{\mu\rho\nu}\left[476 m^2 + 501 m m_{\sigma } + 167 m^2_{\sigma }\right].\label{13}
\end{eqnarray}

\subsection{Results for $T\neq 0$}

The amplitudes for $T\neq 0$ are associated with the same graphs as
in the previous section  but using Eq. (\ref{nzerot}) in the corresponding analytic expressions. With the help of Feynman parametric integrals the 
propagators are combined into a sum of terms with only one factor in the denominator.
The $k_0$ integral is then performed by adequately closing the contour
of integration and applying the residue theorem. Afterwards the $\vec{k}$
integrals are calculated in the static limit. We find

1. $T\not= 0$ part of the graph \ref{Fig2}$b$

\begin{eqnarray}
\left[\Gamma^{3b}_{\mu \nu \rho}
\right]^{T}_{odd} &=&\frac{-112v^2g^5}{3}\sin(p_1\wedge p_2)\nonumber\\
&\times&\left\{-2\int_{0}^{1}dx\int_{T\neq0}\frac{d^{3}k}{(2\pi)^3
}(1-x)x\left[\frac{m^2\epsilon_{\mu\rho\nu}+\epsilon_{\mu\rho\alpha}k^{\alpha}k_{\nu}}{(k^2- \Lambda_1^{2})^3} \right]\right.\nonumber\\ 
&+&\!\left.6m^2\!\int_{0}^{1}\!dx\int_{0}^{1}\!dy\int_{T\neq0}\frac{d^{3}k}{(2\pi)^3}[2-y(1-x)]xy^2\left[\frac{\epsilon_{\mu\alpha\nu}k^{\alpha}k_{\rho}+\epsilon_{\mu\rho\alpha}k^{\alpha}k_{\nu}}{(k^2 - \Lambda_2^{2})^4}\right]\right\},
\label{Grapf2_T}
\end{eqnarray}

\noindent where $\Lambda^2_{1}= m^2_{\sigma} (1-x) + m^2 x $ and $\Lambda^2_{2}= y\Lambda^2_{1}$, whereas, for the graph \ref{Fig2}$c$

\begin{equation}
\left[\Gamma^{3c}_{\mu\nu\rho }\right]^{T}_{odd}=-80v^2g^5\sin(p_1\wedge p_2)
\int_{0}^{1}dx\, x(1+\frac{x}2)\int_{T\neq0}\frac{d^{3}k}{(2\pi)^3}\frac{\epsilon_{\mu\rho\alpha}k^{\alpha}k_{\nu}}{(k^2 - \Lambda_3^{2})^3},
\label{Grapf3_T}
\end{equation}

\noindent where $\Lambda^2_{3}= m^2 (1-x) + m^2_{\sigma} x $.

As a prototype for the calculation of the above expression we consider the
typical integral 

\begin{eqnarray}
I_{\mu\rho\nu}&=&\int_{T\neq0}\frac{d^{3}k}{(2\pi)^{3}}\frac{\epsilon_{\mu\rho\alpha}k^\alpha
k_\nu}{(k^2-\Lambda^2)^3}\nonumber\\
&=&\int_{T\neq0}\frac{d^{3}k}{(2\pi)^{3}}\left[\frac{g_{0\nu}\epsilon_{\mu\rho0}k^{2}_{0}+g_{\nu j}\epsilon_{\mu\rho i}k^i
k^j}{(k^2-\Lambda^2)^3}\right],
\end{eqnarray}

\noindent
where  $\Lambda =\Lambda_i$ with
$i=,1,2,3$.
After integrating over  $k_0$ we obtain

\begin{equation}
I_{\mu\rho\nu}=-2i\int\frac{d^{2}k}{(2\pi)^{2}}[g_{0\nu}\epsilon_{\mu\rho0}F_{1}(\vec{k}^{2})
+g_{\nu i}\epsilon_{\mu\rho i}F_{2}(\vec{k}^{2})],
\end{equation}
\noindent where

\begin{eqnarray}
F_{1}(\vec{k}^{2})&=&\frac{-1+e^{2\beta\omega_{\Lambda}}(-1-\beta\omega_{\Lambda}+\beta^2\omega^2_{\Lambda})+e^{\beta\omega_{\Lambda}}(2+\beta\omega_{\Lambda}+\beta^2\omega^2_{\Lambda})}{16\omega_{\Lambda}^{3}(e^{\beta\omega_{\Lambda}}-1)^3}\nonumber\\ &{\buildrel \beta\rightarrow 0 \over \simeq}&
\frac{1}{32\omega^3_{\Lambda}}+\mathcal{O}(\beta^3)\label{1}
\end{eqnarray}

\noindent and
\begin{eqnarray}
F_{2}(\vec{k}^{2})&=&\frac{3+e^{2\beta\omega_{\Lambda}}(3+3\beta\omega_{\Lambda}+\beta^2\omega^2_{\Lambda})+e^{\beta\omega_{\Lambda}}(-6-3\beta\omega_{\Lambda}+\beta^2\omega^2_{\Lambda})}{16\omega_{\Lambda}^{5}(e^{\beta\omega_{\Lambda}}-1)^3}\nonumber
\\ &{\buildrel \beta\rightarrow 0 \over \simeq}&
\frac{1}{2\omega^6_{\Lambda}\beta}-\frac{3}{32\omega^5_{\Lambda}}+\mathcal{O}(\beta^4)\label{2}
\end{eqnarray}

The expressions in the right hand side of Eqs. (\ref{1}) and (\ref{2}) 
correspond
to the high $T$ limit  of $F_1$ and $F_2$, respectively. 
In this limit the integrals on the spatial components of $k$ are very
simple, giving
\begin{equation}
I_{\mu\rho\nu}=\frac{-i}{16\pi}\left[\frac{1}{\Lambda}\epsilon_{\mu\rho\nu}-\frac{2}{\beta\Lambda^2}g^{i}_{\nu}
\epsilon_{\mu\rho i}\right] + \mathcal{O}(\beta^3).\label{10}
\end{equation}

\noindent
It remains to perform the Feynman parametric integrals which for expressions 
like (\ref{10}) are trivial.

Proceeding as in the example above, the relevant integrals may be
straightforwardly  computed. We list
the results for each contributing graph

\begin{eqnarray}
\left[\Gamma^{\ref{Fig2}b}_{\mu \nu \rho
}\right]^{T}_{odd}&=&\frac{-i v^2 g^5 \sin(p_1 \wedge
  p_2)}{\pi}\left\{\frac{1}{21\beta}\left[A_1(4\epsilon_{\mu\rho\nu}-g^{i}_{\mu}\epsilon_{ i \rho \nu}-g^{i}_{\rho}\epsilon_{\mu i \nu})+A_2g^{i}_{\nu}
\epsilon_{\mu\rho i}\right] \right. \nonumber\\
&+& A_3 \epsilon_{\mu\rho\nu}\Bigg{\}},
\label{fim1}
\end{eqnarray}

\noindent where

\begin{eqnarray}
A_1 &=& 2 \frac{[m^2_{\sigma}m^2 - m^4 + (m^4+m^2m^2_{\sigma})\ln(\frac{m}{m_{\sigma}})]}{(m^2-m^2_{\sigma})^3},\\
A_2&=& \frac{m^4_{\sigma} - m^4 +4 m^2m^2_{\sigma}\ln(\frac{m}{m_{\sigma}})}{(m^2-m^2_{\sigma})^3},\\
A_3&=&\frac{28}{45(m+m_{\sigma})^3}(m^2_{\sigma}-2m^2+3mm_{\sigma}),
\end{eqnarray}

\noindent and

\begin{eqnarray}
\left[\Gamma^{\ref{Fig2}c}_{\mu \nu \rho}\right]^{T}_{odd}&=&
\frac{-80iv^2g^5\sin(p_1 \wedge p_2)}{\pi}\left[\frac{B_1}{\beta}g^{i}_{\nu}\epsilon_{\mu\rho i} + B_2\epsilon_{\mu \nu
    \rho} \right],
\label{fim2}
\end{eqnarray}

\noindent
where

\begin{equation}
B_1=\frac{1}{32(m^2-m^2_{\sigma})^3}\left[5m^4_{\sigma}-12m^2m^2_{\sigma}+7m^4-4(2m^2m^2_{\sigma}-3m^4)\ln(\frac{m_{\sigma}}{m})\right]
\end{equation}

\noindent and

\begin{equation}
B_2=\frac{1}{240(m+m_{\sigma})^3}(13m^2_{\sigma}+39mm_{\sigma} +28m^2).
\end{equation}

Notice that the terms containing $A_3$ and $B_2$ coincide up to a sign with the expressions in
Eqs. (\ref{11}) and (\ref{12}), respectively, so that when computing the
high temperature limit of the three point vertex  function they mutually
cancel.

\section{Conclusions}
\label{sec4}

We can now summarize the results obtained in the previous sections.

(1) Zero temperature:
For small momenta the  corrections to the two and three point
vertex functions given in Eqs. (\ref{ref39}) and (\ref{13}) lead to the
conclusion that the following action is induced (see remark after Eq.
(\ref{ref39}))

\begin{equation}
S^{T=0}_{ind}=\frac{\bar{\kappa}_{1}}{2}\int d^3{x} \epsilon_{\mu \nu \rho}
\left[A^{\mu}\partial^{\nu}A^{\rho}+\frac{2i\bar{g}_{1}}{3}A^{\mu}*A^{\nu}
*A^{\rho}\right],\label{resT2}
\end{equation}

\noindent
where

\begin{equation}
\bar{\kappa}_{1}=\frac{2g^4 v^2}{3\pi}\frac{(2 m +
  m_{\sigma})}{(m+m_{\sigma})^2}
\end{equation}

\noindent
and

\begin{equation} 
\bar{g}_{1}=g\frac{236 m^2+231 m m_{\sigma}+77 m^{2}_{\sigma}}{60 (m+m_{\sigma})(2m+m_{\sigma})}.
\end{equation}

In the limit $m_\sigma=m$ which is relevant for the supersymmetric Chern-Simons-Higgs model \cite{Lee} a great simplification is achieved so
that $\bar{\kappa}_{1}=\frac{g^2}{4\pi}$ and $\bar{g}_{1}=\frac{68}{45}g$.

2) High temperature limit:

\begin{equation}
S^{T}_{ind}\,\,{\buildrel \beta\rightarrow 0 \over \simeq}\,\frac{\bar{\kappa}_{2}}{2 \,\, \beta}\int d^3{x} \epsilon_{0 i j}
\left[A^{0}\partial^{i}A^{j}+\frac{2i\bar{g}_{2}}{3}A^{0}*A^{i}*A^{j}\right],\label{resT3}
\end{equation}

\noindent where

\begin{eqnarray} 
\bar{g}_{2}&=& \frac{g}{168(m^2-m^2_{\sigma})}\nonumber\\
&\times&\left[\frac{721 m^4 - 1248
  m^2m^2_{\sigma}+527 m^4_{\sigma}+(-1248m^4+860 m^2m^2_{\sigma})\ln(\frac{m}{m_{\sigma}})}{m^2_{\sigma}-m^2+2m^2\ln(\frac{m}{m_{\sigma}})}\right]
\end{eqnarray}

\noindent
and

\begin{equation}
\bar{\kappa}_{2}=\frac{2 v^2 g^4}{\pi(m^2-m^2_{\sigma})^2} [m^2_{\sigma}-m^2+2m^2\ln(\frac{m}{m_{\sigma}})].           
\end{equation}

\noindent
Here again a great simplification occurs for equal masses: $\bar{\kappa}_{2}=\frac{1}{4 \pi v^2}$ and $\bar{g}_{2}= \frac{-839g}{252}$.

The leading noncommutative corrections to the two point vertex  function were
also obtained and are given by

\begin{equation}
\Pi^{odd}_{0i}(T=0)=\frac{ g^2 \tilde{p}}{16\pi}(m-\xi)\epsilon_{0ij} p^j,
\label{ref39final}
\end{equation}

\noindent at zero temperature, and

\begin{eqnarray}
{\cal{\pi}}_{0i}^{NC} (p_0=0;T) &=& - \frac{\tilde{p}^2}{8\pi}
\epsilon_{0ij} p^j T \nonumber \\ 
&\times& \left\{ 2 (ve^2)^2 \frac{\partial}{\partial m_{\sigma}^2}\left[
 \frac{(m^4
\log(m/T) - m^4_{\sigma}\log(m_{\sigma}/T))}{m_{\sigma}^2 - m^2}\right]
\nonumber \right. \\
&+& \left. \frac{g^2}{2} \left[2 m \log(m/T) + (m - \xi) F\right]
  \right\} ,\label{resT1}
\end{eqnarray}

\noindent
in the high temperature limit.
From Eqs. (\ref{ref39final})
and (\ref{resT1}) we can see that the commutativity can
be recovered straightforwardly, by considering the limit $\tilde{p}
\rightarrow 0$. In other words, there are no infrared UV/IR singularity appearing
in this limit.
Furthermore, looking at the finite temperature result,
Eq. (\ref{resT1}), in the static limit we can extract the leading
behavior in the high temperature regime and first order correction in
the noncommutative parameter as being proportional to $T
logT$. So, our calculation provides a logarithm correction to
the result obtained in \cite{parityvio} for the commutative version of
the same model. 

The results in Eqs. (\ref{resT2}) and (\ref{resT3}) are formally invariant
under small gauge transformations. Notice however that, because of the
spontaneous breakdown of the symmetry, such property will be certainly
lost whenever other corrections are incorporated. This is already indicated by the form of the noncommutative corrections in Eqs. (\ref{ref39final}) and
(\ref{resT1}).

 It should be pointed out that
previous studies of noncommutative gauge theories have shown that,
as it happens in the commutative non-Abelian CS model,
invariance under large gauge transformation requires that the CS coefficient
be quantized \cite{quantization}. At finite temperature, the 
verification of this property for the effective action is a highly non
trivial task which probably involves all orders of perturbation and
also other (nonlocal) interactions. Even in the
commutative setting, invariance under large non-Abelian gauge
transformation was verified only in simplified situations
\cite{seminara}. In our calculations invariance under large gauge transformation is
certainly partially  broken. In spite of that, our results should still
be a good approximation for small couplings.

\section{Acknowledgments}

This work was partially supported by  Conselho
Nacional de Desenvolvimento Cient\'{\i}fico e Tecnol\'ogico (CNPq) and
 Funda\c{c}\~ao de Amparo \`a Pesquisa do Estado de S\~ao Paulo (FAPESP).

\begin{appendix}
\section{An Useful Integral}
\label{a1}
In this appendix, we will compute a basic integral  that appear in
Eq. (\ref{integral}). Let us define

\begin{eqnarray}
I(m) &\equiv& \int_0^{\infty} k^2 dk J_1 (k \tilde{p})
\frac{\coth{(\beta w_m/2)}}{w_m}. \nonumber \\
&=& \int_0^{\infty} k^2 dk J_1 (k \tilde{p})
\frac{1}{w_m} + 2 \int_0^{\infty} k^2 dk J_1 (k \tilde{p})
\frac{n_B(w_m)}{w_m}.
\end{eqnarray}

Using that 

\begin{equation}
\frac{1}{e^x-1}= \sum_{n=1}^{\infty} e^{-n x}
\end{equation}

\noindent we can rewrite $I(m)$ as

\begin{eqnarray}
I(m)&=& \int_m^{\infty} dx \sqrt{x^2 - m^2} J_1[\tilde{p}\sqrt{x^2 -
m^2}] \nonumber \\
&+& T^2 \sum_0^{\infty} \int_{\beta m}^{\infty} dx \sqrt{x^2 - \beta
m^2} J_1[\tau \sqrt{}x^2 - b^2 m^2] e^{-n x} \nonumber \\
&=&m^2 \int_1^{\infty} dy \sqrt{y^2 -1} J_1[\tilde{p} m \sqrt{y^2 -1}]
\nonumber \\
&+& m^2 \sum_0^{\infty} \int_1^{\infty} dy \sqrt{y^2 - 1} J_1[\tau
\beta m \sqrt{y^2 -1}] e^{-\beta m y n}
\end{eqnarray}

 These integrals can be evaluated using the standard result \cite{gradshteyn}

\begin{equation}
\int_1^{\infty} dx (x^2 -1)^{\nu/2} e^{- \alpha x} J_{\nu}[\beta
\sqrt{x^2 -1}] =\sqrt{\frac{2}{\pi}}\beta^{\nu} (\alpha^2 +
\beta^2)^{-\nu/2  - 1/4} K_{\nu + 1/2} (\sqrt{\alpha^2 + \beta^2}).
\end{equation}

\noindent  Then, it is straightforward to verify that

\begin{eqnarray}
I(m) &=& m^2 e^{- \tilde p  m} \left[\frac{1}{\tilde p  m} + \frac{1}{(\tilde p
 m)^2}\right] \nonumber \\
 &+& \tilde p  m T^2 \sum_{n=1}^{\infty}
\frac{e^{- \beta m\sqrt{n^2 + \tau^2}}}{n^2 +\tau^2} +\tilde  p  T^3
\sum_{n=1}^{\infty} \frac{e^{- \beta m \sqrt{n^2 + \tau^2}}}{(n^2
+\tau^2)^{3/2}}.
\end{eqnarray}

Note that here we have computed both zero temperature and finite
temperature parts together,  although in sections \ref{sec2} and
\ref{sec3} they occur separately.

\end{appendix}

\newpage

\begin{thebibliography}{10}  

\bibitem{dunne_1} Dunne G V 1999  Aspects of Chern
Simons Theory.  Les Houches Summer School in Theoretical Physics: Topological
Aspects of Low Dimensional Physics, Les Houches 1968

\bibitem{coleman} Coleman S and Hill B 1985  \emph{Phys. Lett.} B {\bf 159} 184

\bibitem{jackiw} Jackiw R 2002 \emph{Nucl. Phys. Proc. Suppl.} {\bf 108} 30

\bibitem{witten} Seiberg N and Witten E 1999 \emph{JHEP} {\bf 9909} 032   

\bibitem{seiberg} Minwalla S, Van Raamsdonk M and  Seiberg N 2000
 \emph{JHEP} {\bf 0008} 008 
 
 \bibitem{gomes} Girotti H O, Gomes M, Rivelles V O and  da
 Silva A J 2000 \emph{Nucl. Phys.} B {\bf 587} 299 

  Girotti H O,  Gomes M,
  Petrov A Yu, Rivelles V O and da Silva A J 2001 \emph{Phys. Lett.} B {\bf
 521} 119 

\bibitem{putz} Bichl A, Grimstrup J, Putz V and
   Schweda M 2000 JHEP \textbf{0007} 046.

 Das A and Sheikh-Jabbari M M 2001 \emph{JHEP} {\bf 0106}
 028

Martin C P 2001 \emph{Phy. Lett.} B {\bf 515} 185

\bibitem{fischler} Fischler W, Gommis J, Gorbatov E,
  Kashani-Poor A, Paban S and  Pouliot P 2000 \emph{JHEP} {\bf 05} 024

  Landsteiner K, Lopez E and Tytgat M H G 2000 \emph{JHEP} {\bf 09} 027
 
  Arcioni G and V\'azquez-Mozo M A 2000 \emph{JHEP} {\bf 01} 028

  Brandt F T, Das A, Frenkel J, McKeon D G C and 
  Taylor J C 2002 \emph{Phys. Rev.} D {\bf 66} 045011 

  Chandrasekhar B and Panigrahi P K 2003
 \emph{JHEP} {\bf 0303} 015 
\bibitem{bedaque} Kapusta J I 1979  \emph{Nucl. Phys.} B {\bf
 148} 461 

Arnold P, Vokos S, Bedaque P and Das A 1993
 \emph{Phys. Rev.} D {\bf 47} 4698 

  
 


\bibitem{jack2} Deser S, Jackiw R and  Templeton S  1982 \emph{Annals
   Phys.} {\bf140} 372, 1988 Erratum-ibid. 185:406, 2000 \emph{Annals
   Phys.} {\bf281} 409

   Redlich A 1984 \emph{Phys.Rev.} D {\bf 29} 2366  

   Redlich A  1984 \emph{Phys.Rev.Lett.} {\bf 52} 


\bibitem{pisarski} Pisarski R and Rao S 1985 \emph{Phys. Rev.} D \textbf{32} 2081. 


\bibitem{das1} Brandt F, Das A and Frenkel J 2000
 \emph{Phys. Lett.} B  \textbf{494} 339.

\bibitem{quantization} Nair V P and  Polychronakos A P 2001
 \emph{Phys. Rev. Lett.} {\bf 87} 030403  

 Sheikh-Jabbari M M 2001
 \emph{Phys. Lett.} B {\bf 510} 247 

 Bak D, Lee K and Park J H 2001
 \emph{Phys. Rev. Lett.} {\bf 87} 030402  

 Chen Guang-Hong and
  Wu Yong-Shi 2001 \emph{Nucl. Phys.} B {\bf 593} 562 

\bibitem{khlebi} Khlebinikov S and Shaposhnikov M 1991 \emph{Phys Lett} B
  \textbf{254} 148

 Khare A, MacKenzie R, Paranjape M and Panigrahi P 1995 \emph{Phys. Lett.} B
 \textbf{355} 236

Khare A, MacKenzie R and Paranjape M 1995 \emph{Phys. Lett.} B \textbf{343}
239

Chen L, Dunne G, Haller K and Lim-Lombridas E 1995
\emph{Phys. Lett.} B \textbf{348} 468.

 Kao H 1998  \emph{Phys. Rev.} D \textbf{57} 7416

\bibitem{frenkel} Brandt F T, Frenkel J and  McKeon D G C 2002
 \emph{Phys. Rev.} D {\bf 65} 125029 


\bibitem{parityvio} Alves V S, Das A, Dunne G V and
  Perez S 2002 \emph{Phys. Rev.} D {\bf 65} 085011

\bibitem{gronewold} Gronewold H J 1946 \emph{Physica} (Amsterdan) {\bf 12}
 405 
 
 Weyl H 1949 \emph{Z. Physik} {\bf 46} 1  

 Moyal J E 1949
 \emph{Proc. Cambridge Philos. Soc.} {\bf 45}  99

 Doplicher S, Fredenhagem K and Roberts J E 1995
 \emph{Commun. Math. Phys.} {\bf 172} 187

\bibitem{douglas} Douglas M R and Nekrasov N A 2001 \emph{Rev. Mod. Phys.}
 {\bf 73} 977  

 Szabo R J 2003 \emph{Phys. Rept.} {\bf 378} 207

 Gomes M 2002  V  in Proceedings of the XI Jorge Andre Swieca Summer
 School, edited by Alves G, Eboli O and Rivelles V, World Scientific

Girotti H 2003 \emph{Noncommutative Quantum Field Theory} {\bf hep-th}/0301237 

\bibitem{leon} Lewin L 1958 \emph{Dilogarithms and associated functions}
(London: Macdonald)

\bibitem{kapusta} Kapusta J I 1993 \emph{Finite-temperature field Theory}
(Cambridge)
\bibitem{Lee} Kao H-C, Lee K, Lee C,  and Lee T 1994 \emph{Phys. Lett.} B {\bf 341} 181

\bibitem{seminara} Dunne G, Lee K and  Lu C 1997 \emph{Phys. Rev. Lett.} {\bf
  78} 3434 

Deser S, Griguolo L and Seminara D 1997
  \emph{Phys. Rev. Lett.} {\bf 79} 1976  

Deser S, Griguolo L and Seminara D 1998 \emph{Phys. Rev.} D {\bf 57} 7444
  

 Deser S, Griguolo L and Seminara D 1998 \emph{Commun. Math. Phys.} {\bf 197} 443 

 Deser S, Griguolo L and Seminara D 2003 Phys Rev D {\bf 67} 065016

 Fosco C D, Rossini G L and Schaposnik F A 1997
 \emph{Phys. Rev. Lett.} {\bf 79} 1980, 


 Fosco C D, Rossini G L and Schaposnik F A 1997
 \emph{Phys. Rev. Lett.} {\bf 79} 4296(E)  

 Brandt F T, Das A and  
Frenkel J 2002  \emph{Phys. Rev.} D {\bf 65} 065013 

 Brandt F T, Das A,
 Frenkel J,  Pereira S and Taylor J C 2001  \emph{Phys. Rev.} D {\bf 64} 065018 



 \bibitem{gradshteyn} Gradshteyn I S and  Ryzhik M 1980  Table of
 Integral, Series and Products Academic Press New York


\end{thebibliography}
\end{document}